\documentclass[aps,prl,twocolumn,showpacs,amsmath,amssymb,superscriptaddress]{revtex4}

\usepackage{graphicx}
\usepackage{dcolumn}
\usepackage{bm}

\def\beq{\begin{equation}}
\def\eeq{\end{equation}}
\def\bea{\begin{eqnarray}}
\def\eea{\end{eqnarray}}

\begin{document}

\title{Unconventional magnetism via optical pumping of interacting spin systems}

\author{Tony E. Lee}
\affiliation
{ITAMP, Harvard-Smithsonian Center for Astrophysics, Cambridge, MA 02138, USA}
\author{Sarang Gopalakrishnan}
\affiliation{Department of Physics, Harvard University, Cambridge, MA 02138, USA}
\author{Mikhail D. Lukin}
\affiliation{Department of Physics, Harvard University, Cambridge, MA 02138, USA}

\date{\today}

\begin{abstract}
We consider strongly interacting systems of effective spins, subject to dissipative spin-flip processes associated with optical pumping. We predict the existence of novel magnetic phases in the steady-state of this system, which emerge due to the competition between coherent and dissipative processes. Specifically, for strongly anisotropic spin-spin interactions, we find ferromagnetic, antiferromagnetic, spin-density-wave, and staggered-XY steady states, which are separated by nonequilibrium phase transitions meeting at a Lifshitz point. These transitions are accompanied by quantum correlations, resulting in spin squeezing. Experimental implementations in ultracold atoms  and trapped ions are discussed. 
\end{abstract}

\pacs{}
\maketitle

Exotic magnetic states play a central role in the physics of quantum  many-body systems, and have been explored in  a wide variety of strongly correlated materials \cite{dagotto94}. Realizing and exploring magnetic states has recently emerged as  a central goal in ultracold atomic physics \cite{duan03,bloch08}. Due to highly controllable and tunable interactions, ensembles of ultracold neutral atoms and ions may provide a unique 
laboratory to study exotic quantum magnetism \cite{duan03,bloch08,kim10,simon11,sg:frust,strack2011,gorshkov11,britton12}
Among the main obstacles are relatively small energy scales 
associated with magnetic  ordering (e.g., the superexchange scale in the Hubbard model), requiring cooling atomic systems down to very low temperatures \cite{duan03} and  the slow timescales involved in spin thermalization \cite{capogrosso10,medley11,mckay11}. Furthermore, 
ultracold atoms are fundamentally open, driven quantum systems far away from their absolute thermal equilibrium. This motivates the exploration of spin dynamics in the presence of driving and dissipation~\cite{diehl2008, verstraete2009,  diehl10, kasprzak2006, carusotto2013, dallatorre2010, weimer10, cho2011, lee11, lee12,  ates12, olmos12, nissen12, kessler2012, hoening2012, fossfeig13, dallatorre2013, carr13}.

Recently a number of  schemes involving dissipation to create magnetic phases have been proposed. These typically use engineered reservoirs involving coupling multiple lattice sites~\cite{diehl2008, verstraete2009, diehl10}. At the same time, one expects single-site dissipation such as spontaneous decay to be detrimental to realizing interesting magnetic states, resulting e.g. in unwanted decoherence.  In this Letter, we demonstrate that 
optical pumping and spontaneous decay can instead \emph{enrich} the phase diagram, resulting in new phases and phase transitions that do not exist in conventional equilibrium systems. Significantly, these novel states can be observed under conditions when realization of conventional, equilibrium states is difficult.  

The key idea of this work can be understood by considering the anisotropic spin-$1/2$ Heisenberg model (i.e., the XYZ model), which  is governed by the Hamiltonian 

\begin{eqnarray}
H&=&\frac{1}{2d}\sum_{\langle mn\rangle} (J_x\sigma^x_m\sigma^x_n + J_y\sigma^y_m\sigma^y_n + J_z\sigma^z_m\sigma^z_n), \label{eq:H}
\end{eqnarray}
where $\sigma^x_n,\sigma^y_n,\sigma^z_n$ are the Pauli matrices for an effective spin $n$. We assume that the spins are localized on a $d$-dimensional cubic lattice with nearest-neighbor interactions. In the presence of conventional optical pumping, this Hamiltonian is augmented with a dissipative process that flips the spins down at some rate $\gamma$ (i.e., it corresponds to the jump operator $\sigma^-_n$ on every site, where $\sigma^\pm_n=(\sigma^x_n\pm i\sigma^y_n)/2$).

\begin{figure}[b]
\centering
\includegraphics[width=3 in,clip]{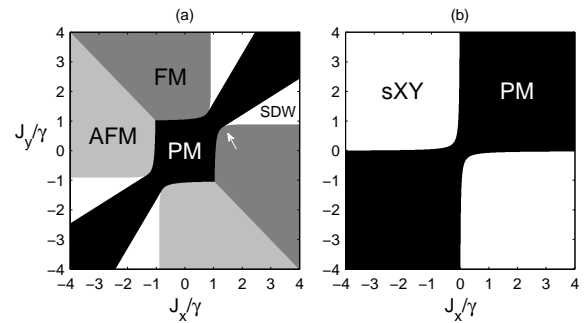}
\caption{\label{fig:phasediagram}Mean-field phase diagrams for the dissipative XYZ model with (a) $J_z/\gamma=1$ and (b) $J_z=0$, showing the different phases: paramagnetic (PM), ferromagnetic (FM), antiferromagnetic (AFM), spin-density-wave (SDW), and staggered-XY (sXY). The white arrow points to a Lifshitz point.}
\end{figure}

The steady state of this open many-body system is easy to understand in the case of isotropic spin-spin interactions, namely the XXZ model (with either ferromagnetic or antiferromagnetic couplings). 
For this, the Hamiltonian can be rewritten in the form $H = (1/2d)\sum [2J_x (\sigma_m^+ \sigma_n^- + \sigma_m^-\sigma_n^+) + J_z \sigma^z_m \sigma^z_n]$. This Hamiltonian conserves the total number of spins in the $|\uparrow\rangle$ state, and therefore does nothing to counteract the spontaneous decay. Thus, the steady state is a trivial dark state with all spins polarized, ${|\downarrow\downarrow\cdots\downarrow\rangle\langle\downarrow\downarrow\cdots\downarrow|}$, so the XXZ model never experiences a phase transition in the presence of dissipation, regardless of $J_x$ and $J_z$.

\begin{figure}[b]
\centering
\includegraphics[width=2.5 in]{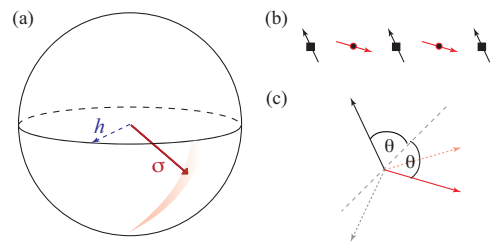}
\caption{\label{fig:blochsphere}(a) Bloch-sphere plot, showing mean-field values of $\langle\vec{\sigma}\rangle$ (solid red arrow) and effective magnetic field (dashed blue arrow) for $J_x/\gamma=-J_y/\gamma=1$, $J_z=0$. The vectors are normalized to unit length. (b), (c) show the sXY phase in the $xy$ plane of the Bloch sphere. (b) shows one possible stable configuration. Black and red arrows correspond to  sublattices A and B. (c) shows that the A sublattice (black solid arrow) generates a magnetic field (black dashed arrow) that the B sublattice (red solid arrow) precesses around. Similarly, the B sublattice generates a magnetic field (red dashed arrow) that the A sublattice precesses around. The angle $\theta$ can take any value.}
\end{figure}

However, new types of magnetic order emerge  for strongly anisotropic couplings. The crucial role of anisotropy can be understood as follows. Each spin experiences an effective magnetic field $(J_x\langle\sigma^x\rangle,J_y\langle\sigma^y\rangle,J_z\langle\sigma^z\rangle)$, which depends on the direction of its neighbors  [Fig.~\ref{fig:blochsphere}(a)]. It precesses about this effective field and also decays towards $|\downarrow\rangle$. In order for the spin to point away from $|\downarrow\rangle$ in steady state, its precession must be strong enough to counteract the decay. In the isotropic case, the spin is always parallel to the magnetic field, so there is no precession at all. On the other hand, when the couplings are sufficiently anisotropic (e.g., $J_x \approx - J_y$), the spin is roughly \emph{perpendicular }to the magnetic field, so the precession is strong enough to point the spin away from $|\downarrow\rangle$ [Fig.~\ref{fig:blochsphere}(a)]. This is in sharp contrast with thermal equilibrium state, in which the spin tries to align with the magnetic field rather than precess about it. 

This competition between precessional and dissipative dynamics gives rise to a remarkable phase diagram (Fig.~\ref{fig:phasediagram}), including ferromagnetic and antiferromagnetic phases as well as spin-density-wave and staggered-XY phases that do not exist in equilibrium. 
The spin-density-wave, paramagnetic, and ferromagnetic phases meet at multicritical Lifshitz points, at which the period of the spin-density wave diverges~\cite{hornreich1975}; such Lifshitz points have been seen in equilibrium magnets with long-range interactions~\cite{grousson2000, jamei2005}, 
but generally do not exist in nearest-neighbor spin models. 
In addition, we find that a continuous symmetry emerges for certain couplings; the spontaneous breaking of this symmetry leads to a phase we call the staggered-XY phase. 
Finally, we find that quantum correlations (as measured by spin squeezing) persist near the phase transitions.

The model described here can be implemented in systems of trapped ions or systems of ultracold atoms with anisotropic superexchange or dipolar interactions. The spin states $|\uparrow\rangle$ and $|\downarrow\rangle$ correspond to two electronic states of the ion or atom. In the case of ions, the spin-spin interaction is obtained through virtual transitions involving motional sidebands~\cite{kim10, molmer99, porras04}. In the case of ultracold atoms, the spin-spin interaction is obtained using a two-photon resonance that excites and de-excites atoms in pairs \cite{bouchoule02}, as explained in the Supplementary Material, or using superexchange interactions in $p$-band optical lattices \cite{pinheiro13}. In all cases, dissipation can be controllably introduced using optical pumping. 

\emph{Model.}
We now turn to detailed analysis of the phenomena outlined above. The dynamics of the many-body system are given by a master equation for the density matrix $\rho$:
\begin{eqnarray}
\dot{\rho}&=&-i[H,\rho] + \gamma\sum_n\left[\sigma^-_n\rho\sigma^+_n -\frac{1}{2}(\sigma^+_n\sigma^-_n\rho + \rho\sigma^+_n\sigma^-_n)\right].\nonumber\\ \label{eq:master}
\end{eqnarray}
Equation \eqref{eq:master} has a unique steady state solution \cite{schirmer10}, and we are interested in whether the steady state exhibits a phase transition as the parameters $J_x,J_y,J_z$ change. Note that the decay is independent for each spin, in contrast with the Dicke model~\cite{nagy2011, dallatorre2013}. Furthermore, the spins are not in equilibrium with the environmental bath. Thus, in contrast with the spin-boson model~\cite{leggett:rmp,werner05}, the steady state is \textit{not} the joint ground state of the system and environment.

The master equation has a $Z_2$ symmetry ($\sigma^x_n,\sigma^y_n\rightarrow-\sigma^x_n,-\sigma^y_n$), which is spontaneously broken in the ordered phases. In practice, there may also be dephasing noise, leading to dissipative terms in Eq.~\eqref{eq:master} like $\sigma^z_n\rho\sigma^z_n$; since the $Z_2$ symmetry is unaffected by these terms, the phase transitions we describe are robust to dephasing, although the phase boundaries are shifted.

\emph{Mean-field theory.}
We begin by solving for the steady states of the model Eq.~\eqref{eq:master} at the level of mean-field theory. We allow the mean field to vary on each site to account for spatially inhomogeneous states \cite{lee11}. The mean-field equations, which are simply nonlinear Bloch equations, are:
\begin{eqnarray}
\frac{d\langle\sigma_n^x\rangle}{dt}&=&-\frac{\gamma}{2}\langle\sigma_n^x\rangle+\frac{1}{d}\sum_m[J_y\langle\sigma_n^z\rangle\langle\sigma_{m}^y\rangle-J_z\langle\sigma_n^y\rangle\langle\sigma_{m}^z\rangle],\nonumber\\
\frac{d\langle\sigma_n^y\rangle}{dt}&=&-\frac{\gamma}{2}\langle\sigma_n^y\rangle+\frac{1}{d}\sum_m[J_z\langle\sigma_n^x\rangle\langle\sigma_{m}^z\rangle-J_x\langle\sigma_n^z\rangle\langle\sigma_{m}^x\rangle],\nonumber\\
\frac{d\langle\sigma_n^z\rangle}{dt}&=&-\gamma(\langle\sigma_n^z\rangle+1)\nonumber\\ &&\quad\quad+\frac{1}{d}\sum_m[J_x\langle\sigma_n^y\rangle\langle\sigma_{m}^x\rangle-J_y\langle\sigma_n^x\rangle\langle\sigma_{m}^y\rangle],
\end{eqnarray}
where the sum over $m$ is taken over nearest neighbors of $n$. (A related model with only dephasing noise was studied in Ref.~\cite{anglin01a,anglin01b}. Another related model with an external field and nonlinear damping was studied using the Landau-Lifshitz-Gilbert equation \cite{gilbert04,lakshmanan11}.) 

Clearly, there is always a fixed-point solution, $\langle\sigma^x_n\rangle=\langle\sigma^y_n\rangle=0,\langle\sigma^z_n\rangle=-1$, in which all the spins are pointing down. We call this the paramagnetic (PM) phase, since it does not break the $Z_2$ symmetry of Eq.~\eqref{eq:master}. We now consider the linear stability of the PM phase as a function of $J_x,J_y,J_z$ \cite{cross93}. We consider $d$-dimensional perturbations with wave vector $\vec{k}=(k_1,k_2,\ldots,k_d)$ where $k_{\ell}=2\pi/a_{\ell}$ and $a_{\ell}$ is an integer. We find that the PM phase is unstable to perturbations of wave vector $\vec{k}$ when
\begin{eqnarray}
\left(\frac{J_x}{d}\sum_{\ell=1}^d\cos k_{\ell} - J_z\right)\left(\frac{J_y}{d} \sum_{\ell=1}^d\cos k_{\ell} - J_z\right) &<&-\frac{\gamma^2}{16}.\quad\label{eq:condition}
\end{eqnarray}
This condition is satisfied only when the couplings are sufficiently anisotropic.

When the PM phase is unstable, the system ends up in a time-independent steady state with $\langle\sigma_n^x\rangle,\langle\sigma_n^y\rangle\neq0$, so it breaks the $Z_2$ symmetry of the master equation. There are four types of ordered phases: (i) Spatially uniform state, which we call the ferromagnetic (FM) phase, resulting from instability of the PM phase to $k_{\ell}=0$ for all $\ell$. (ii) Spatially modulated state with a period of two lattice sites in all directions, i.e., the system divides into two sublattices. We call this the antiferromagnetic (AFM) phase, and it results from instability to $k_{\ell}=\pi$ for all $\ell$. (iii) Spatially modulated state with a period greater than two lattice sites in at least one direction, which we call the spin-density-wave (SDW) phase. This results from instability to all other $k_{\ell}$. (iv) When $J_z=0$, there is also a staggered-XY (sXY) phase, resulting from instability to both $k_{\ell}=0,\pi$, which is discussed below. The phase diagram is shown in Fig.~\ref{fig:phasediagram}(a)--(b). The transitions from the PM phase are continuous, while the FM-AFM transition is discontinuous.

\begin{figure}[t]
\centering
\includegraphics[width=3 in,clip]{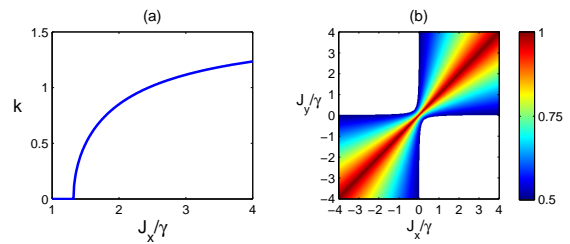}
\caption{\label{fig:lifshitz_squeezing} (a) Unstable wave vector $k$ along the lower boundary of the PM phase in Fig.~\ref{fig:phasediagram}(a). A Lifshitz point occurs at $J_x/\gamma=1.32$. For convenience, only one-dimensional wave vectors are shown. (b) Squeezing parameter $\zeta^2$, calculated in the Gaussian approximation for $J_z=0$. The sXY phase has been whited-out, since the Gaussian approximation is not valid there.}
\end{figure}

We note two unusual features of this phase diagram. First, along the boundary between the PM and SDW phases, the $\vec{k}$ at which the instability of the PM occurs approaches 0, meaning that the period of the SDW diverges [Fig.~\ref{fig:lifshitz_squeezing}(a)]. This line culminates in a multicritical Lifshitz point \cite{hornreich1975} between the PM, FM, and SDW phases. Lifshitz points occur in magnetic models with competing interactions~\cite{grousson2000, jamei2005}, but are not found in equilibrium nearest-neighbor magnets: thus, their existence in nearest-neighbor magnets out of equilibrium indicates that nonequilibrium phase diagrams can be qualitatively richer than those in equilibrium. Lifshitz points show enhanced fluctuation effects relative to conventional critical points~\cite{hornreich1975}, and hence offer a rich venue for studying quantum fluctuations away from equilibrium. 

The second distinctive feature of the phase diagram is that the ordered phase breaks a \emph{continuous} symmetry when $J_z=0$. In this case, the system divides into two sublattices like in the AFM phase. However, the angle between the two sublattices can take any value. In the specific case of $J_x=-J_y$, the spins on the A and B sublattices are at angles $\theta$ and $-\theta$ relative to the $x=y$ line on the Bloch sphere [Fig.~\ref{fig:blochsphere}(b)]. Any value of $\theta$ corresponds to a stable configuration, since the sublattices remain perpendicular to each other's magnetic field [Fig.~\ref{fig:blochsphere}(c)]. Upon ordering, this continuous $U(1)$ symmetry between the sublattice spin orientations is spontaneously broken, leading to a phase we call the staggered-XY (sXY) phase. This phase has vortex-like topological defects around which the relative orientation between A- and B-sublattice spins rotates by $2\pi$.

\emph{Comparison with equilibrium.} It is instructive to contrast the above results with the equilibrium case (for $d > 1$). The equilibrium ground state of Eq.~\eqref{eq:H} is ordered for any $J_x,J_y,J_z$~\cite{sachdev:book}. The magnetization axis is determined by the strongest of the coupling constants, and the sign of that coupling determines whether the ordering is ferromagnetic or antiferromagnetic. 
%
%
%
%
Evidently, the nonequilibrium phase diagram exhibits qualitatively different behavior from this equilibrium case. The qualitative differences between equilibrium and nonequilibrium remain even in the limit $\gamma \rightarrow 0$, although the steady state takes an increasingly long time to reach.

\emph{Fluctuation effects}. We now turn from mean-field theory to an analysis of fluctuations. Such an analysis was recently performed for driven polariton condensates~\cite{sieberer2013} and suggests that the \emph{static} critical properties (i.e., renormalization-group fixed points) of a driven Markovian system are related to finite-temperature equilibrium critical properties. This would indicate that the dissipative XYZ model discussed here undergoes true phase transitions in two or more dimensions. 



We estimate fluctuation effects and squeezing in the Gaussian approximation by mapping the spins to hardcore bosons~\cite{sachdev:book}: $\sigma^+_n \rightarrow b^\dagger_n, \sigma^z_n \rightarrow 2 b^\dagger_n  b_n - 1$. This gives a reliable approximation in the PM phase, where $\langle\sigma^z_n\rangle \approx -1$. To Gaussian order (which includes relaxing the hardcore constraint), the resulting Hamiltonian is
\bea
H & = & \frac{1}{2d} \Big[ (J_x + J_y) \sum_{\langle mn \rangle} (b^\dagger_m b_n +b_m b^\dagger_n) \\ &&   + (J_x - J_y) \sum_{\langle mn \rangle} (b^\dagger_m b^\dagger_n + b_m b_n) - 4d J_z \sum_n b^\dagger_n b_n \Big], \nonumber
\eea
and the dissipative terms in the master equation are $\gamma\sum_n[b_n \rho b^\dagger_n-\frac{1}{2}(b^\dagger_n b_n\rho + \rho b^\dagger_n b_n)]$. We now use standard Keldysh path-integral techniques~\cite{kamenev:review} to compute the relaxation rate, $\langle \sigma^z \rangle$, and the squeezing. We summarize the results here and provide details in the Supplemental Material.

(1) \emph{Relaxation rate}. The rate at which the steady state is approached can be read off from the poles of the retarded Green's function. For notational simplicity, we assume $d=1$ here.  In the Gaussian approximation, the lowest pole has complex frequency $- i \gamma/2\pm 2\sqrt{(J_x \cos k - J_z) (J_y \cos k - J_z)} $. A continuous phase transition occurs when the frequency of this pole approaches zero; this precisely recovers Eq.~\eqref{eq:condition}.

(2) \emph{Below-threshold fluctuations}. Near the transition, one expects to find nonanalytic behavior in the number of up spins, $\sum_n\langle\sigma^z_n\rangle$.
For $J_z = 0$, this scales as $\langle \sigma_z \rangle \sim (\gamma^2 + 16 J_x J_y)^{(d-2)/2}$. The divergence for $d=1$ renders the Gaussian approximation inconsistent, and is related, as we shall show in a future work, to the absence of a phase transition in one dimension (consistent with the polariton-BEC case~\cite{sieberer2013}). 

(3) \emph{Squeezing}. We find that spin squeezing, a measure of quantum correlations, persists near the transition. It can be calculated using the definition of squeezing for bosons \cite{ma2011}: $\zeta^2 =1 + 2 (\langle b^\dagger b \rangle - |\langle b\rangle|^2) - 2 | \langle b ^2 \rangle-\langle b\rangle^2|$. For the case of $J_z=0$, as the phase boundary is approached, $\zeta^2 \rightarrow \frac{1}{2}$ in the thermodynamic limit for the $k = 0, \pi$ modes, signaling the presence of quantum correlations [Fig.~\ref{fig:lifshitz_squeezing}(b)].

\begin{figure}[t]
\centering
\includegraphics[width=3 in,clip]{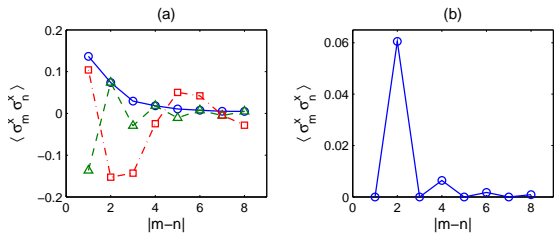}
\caption{\label{fig:correlations} Correlation function $\langle\sigma^x_m\sigma^x_n\rangle$ for 1D chain of 16 spins, from simulating the master equation. (a) $J_z/\gamma=1$, showing remnant of FM for $J_x/\gamma=2$, $J_y=0$ (blue circles, solid line); remnant of AFM for $J_x/\gamma=-2$, $J_y=0$ (green triangles, dashed line); remnant of SDW for $J_x/\gamma=4$, $J_y/\gamma=2$ (red squares, dash-dotted line). The period of the SDW matches the mean-field prediction (5.3 sites). (b) $J_x/\gamma=-J_y/\gamma=1,J_z/\gamma=0$, showing remnant of sXY phase.}
\end{figure}


\emph{Comparison with numerics.} 
We have also simulated the Eq.~\eqref{eq:master} in 1D using the method of quantum trajectories \cite{dalibard92}. Although there is presumably no phase transition in 1D, the numerical results already show qualitative features predicted by mean-field theory. For example, when mean-field theory predicts FM, the correlation $\langle\sigma^x_m\sigma^x_n\rangle$ is positive for all distances [Fig.~\ref{fig:correlations}(a)]. When there should be AFM, the correlation alternates sign. When there should be SDW, the correlation varies with a wavelength that matches the mean-field value. When there should be sXY, $\langle\sigma^x_m\sigma^x_n\rangle$ and $\langle\sigma^y_m\sigma^y_n\rangle$ are both 0 for odd distances and positive for even distances [Fig.~\ref{fig:correlations}(b)]. Furthermore, the gap of the Liouvillian approaches 0 at the boundary of the PM phase, consistent with the Gaussian approximation (see Supplemental Material).

\emph{Experimental realization.} The dissipative XYZ model can be implemented experimentally using trapped ions. One can use ${}^{171}\text{Yb}^+$ and let $|\downarrow\rangle$ and $|\uparrow\rangle$ correspond to ${}^2S_{1/2}|F=0,m_F=0\rangle$ and ${}^2D_{3/2}|F=2,m_F=0\rangle$. In the presence of laser beams judiciously detuned from certain motional sidebands, the ions interact via Eq.~\eqref{eq:H} \cite{kim10, molmer99, porras04}. $J_x,J_y,J_z$ can be on the order of 1-5 kHz, and their magnitudes and signs can be varied by changing the laser detunings \cite{kim10}. By admixing a small component ($10^{-4}$) of ${}^2P_{3/2}$ using an off-resonant laser, one broadens the linewidth of $|\uparrow\rangle$ to 2 kHz. (To make this a closed cycle, additional lasers optically pump back into $|\downarrow\rangle$ on a much faster timescale.) Thus, the parameter space shown in Fig.~\ref{fig:phasediagram} is experimentally achievable. This setup can implement an arbitrary lattice topology for a large number of ions \cite{korenbilt12,britton12}. 

A variety of other realizations of the XYZ model are also possible. One approach is to use ultracold atoms coupled via dipole-dipole interactions.  The XYZ Hamiltonian is implemented by driving a two-photon resonance, so that atoms are excited and de-excited in pairs, as explained in the Supplemental Material. This scheme can be realized using  Rydberg-dressed atoms \cite{henkel10}, Rydberg atoms \cite{lukin01,saffman10,bouchoule02}, or dipolar atoms or molecules \cite{lahaye09}. We show explicitly in the Supplemental Material that, for Rydberg-dressed atoms, the parameters needed for the phase transitions (Fig.~\ref{fig:phasediagram}) are experimentally achievable. %
%
%
Finally, one can adapt a recent proposal for realizing XYZ models via superexchange in $p$-band optical lattices~\cite{pinheiro13} to include dissipation, by optically pumping the atoms into the $p_x$ orbital via an intermediate excited orbital (e.g., $d_{x^2 - y^2}$) that does not decay into the $s$ band.

\emph{Conclusion.} In summary, we have computed the phase diagram of anisotropic spin models subject to spontaneous decay, and shown that these models exhibit phases (SDW and sXY) and phase transitions (Lifshitz point) that are not found in similar equilibrium models. The qualitative differences can be traced to the fact that in equilibrium, spins align with the magnetic field, whereas away from equilibrium, they precess about it. We find that quantum correlations, as measured by squeezing, persist near the dissipative transitions. This work paves the way for future explorations of critical behavior and nonequilibrium fluctuations near the phase transitions we have identified. A particularly intriguing question is how frustrated interactions (due to a triangular lattice) affect the AFM and sXY phases.

We thank Philipp Strack, Eric Kessler, Chris Laumann, Norman Yao, Hendrik Weimer, and Rajibul Islam for useful discussions. This work was supported by NSF through a grant to ITAMP, the Harvard Quantum Optics Center, the Center for Ultracold Atoms, and DARPA.

\bibliography{xyzmodel}

\begin{widetext}
\appendix


\begin{center}
\large{\bf Supplemental material:}\\
\large{\bf Unconventional magnetism via optical pumping of interacting spin systems}\\ 
\vspace{12pt}
\small{Tony E. Lee, Sarang Gopalakrishnan, and Mikhail D. Lukin}
\end{center}

\vspace{12pt}

In this supplement, we first discuss a level scheme for experimentally realizing the XYZ Hamiltonian using dipole-dipole interactions. Then we elaborate on the calculation of excitation gaps and fluctuations presented in the main text. Finally, we show numerical results for the Liouvillian gap.

\section{Level scheme for anisotropic spin Hamiltonians}

We outline a general scheme for realizing the anisotropic Hamiltonian $J_x\sigma^x_m\sigma^x_n + J_y\sigma^y_m\sigma^y_n + J_z\sigma^z_m\sigma^z_n$ starting with atoms with Ising interactions, which can originate from dipole-dipole interactions. (Our scheme is similar to that of Ref.~\cite{bouchoule02}, but we use two-photon transitions instead of four-photon transitions.) This desired Hamiltonian can be rewritten in terms of raising and lowering operators as $J_{ff}(\sigma^+_m \sigma^-_n + \sigma^-_m \sigma^+_n) + J_{sq}(\sigma^+_m \sigma^+_n + \sigma^-_m \sigma^-_n) + J_z\sigma^z_m\sigma^z_n$, where $J_{ff}=J_x+J_y$ and $J_{sq}=J_x-J_y$. The first term corresponds to isotropic, flip-flop interactions; the second corresponds to anisotropic, ``squeezing'' interactions that raise or lower a pair of neighboring spins at once. Our primary interest is in realizing the regime where $J_x \approx -J_y$, so that the flip-flop interaction is much weaker than the squeezing interaction.

\begin{figure}[ht]
\begin{center}
\includegraphics{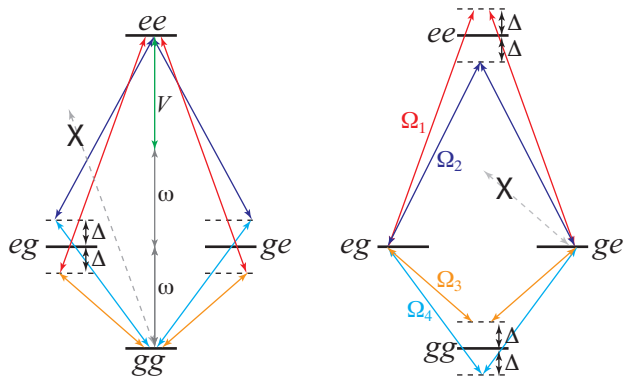}
\caption{Level structure and laser coupling scheme for realizing anisotropic XYZ spin models. We assume two-level atoms, with a transition frequency $\omega$, and an interaction of strength $V$ in the $z$ basis. The levels are coupled using four lasers $\Omega_1$ (red), $\Omega_2$ (dark blue), $\Omega_3$ (orange) and $\Omega_4$ (light blue), such that the pairs $(\Omega_1, \Omega_3)$ and $(\Omega_2, \Omega_4)$ are on two-photon resonance. We take the detunings $\Delta$ to be much smaller than $V$. The left panel shows processes that involve raising or lowering two spins at a time; the right panel shows flip-flop processes. The processes denoted with gray arrows and marked with an X are suppressed because the associated energy denominators are large.}
\label{levels}
\end{center}
\end{figure}

Consider two-level atoms, each with ground state $|g\rangle$ and excited state $|e\rangle$. We explain our scheme assuming that the dipole-dipole interaction exists only between the excited states (like with Rydberg atoms), but the scheme also works when there are interactions between ground states or between ground and excited states. The bare atomic Hamiltonian for two atoms takes the form 
\beq
H = \omega (\sigma^z_1 + \sigma^z_2) + V \sigma^{ee}_1 \sigma^{ee}_2,\label{eq:bareH}
\eeq
where $\omega$ is the transition frequency from $g$ to $e$, $\sigma^{ee}_m\equiv |e\rangle\langle e|_m$, and $V$ is the level shift due to the interaction between the excited states. We now couple the atoms using four lasers $\Omega_1,\Omega_2,\Omega_3,\Omega_4$ as shown in Fig.~\ref{levels}, such that pairs of lasers $(\Omega_1, \Omega_3)$ and $(\Omega_2, \Omega_4)$ are on two-photon resonance, and their detunings from the intermediate atomic levels are $\Delta$. Using the assumption that $\Delta \ll V$, we ignore intermediate states with energy denominators $\sim V$. Then the allowed transitions are those shown in Fig.~\ref{levels}; from second-order perturbation theory,
\bea
J_{ff} & = & -\frac{\Omega_1^2}{2\Delta} + \frac{\Omega_2^2}{2\Delta} - \frac{\Omega_3^2}{2\Delta} + \frac{\Omega_4^2}{2\Delta} \\
J_{sq} & = & \frac{\Omega_1 \Omega_3}{\Delta} - \frac{\Omega_2 \Omega_4}{\Delta}
\eea
Note that in the limit of $\Delta\ll V$, $J_{ff}$ and $J_{sq}$ are independent of $V$.
It follows, in particular, that if we choose all four $\Omega_i$ to have the same magnitude, and arrange the relative phases so that $\Omega_1 \Omega_3 = - \Omega_2 \Omega_4$, then the flip-flop term cancel out completely, but the squeezing terms from the two pairs of lasers add constructively: $J_{ff}=0, J_{sq}=2\Omega_1^2/\Delta$ and thus $J_x=-J_y=\Omega_1^2/\Delta$. To obtain the $J_z\sigma^z_m\sigma^z_n$ term in the Hamiltonian, one detunes the lasers from two-photon resonance.

The Ising Hamiltonian Eq.~(\ref{eq:bareH}) and thus the two-photon resonance scheme can be realized in a number of different settings: Rydberg atoms \cite{lukin01,saffman10, bouchoule02}, Rydberg-dressed atoms \cite{henkel10}, or dipolar atoms or molecules \cite{lahaye09}. In the Rydberg case, the interactions are naturally Ising in character. Using dipolar molecules, one can realize Ising interactions by using the $m = \pm 1$ states in the $J = 1$ rotational manifold~\cite{yao12}. These levels are separated from the $m = 0$ state by an electric field-induced splitting; within the $m = \pm 1$ manifold, flip-flop interactions are prevented by angular momentum conservation, so that the dipolar interaction is purely Ising. 

In each of the cases mentioned above, one can tunably engineer dissipation by dressing one of the spin states off-resonantly with a short-lived excited state. For concreteness, we provide example numbers for the Rydberg-dressed case. Let $|\downarrow\rangle$ and $|\uparrow\rangle$ correspond to the $F=1$ and $F=2$ hyperfine levels of the $5S_{1/2}$ ground state  of ${}^{87}\text{Rb}$. By dressing $|\uparrow\rangle$ with a small component ($10^{-2}$) of the 50S Rydberg state, the dipole-dipole interaction at a distance of 500 nm is 44 MHz \cite{reinhard07}. By doing the two-photon scheme with $\Omega_i=\text{1 MHz}$ and $\Delta=\text{10 MHz}$, one obtains $J_x=-J_y=\text{100 kHz}$. Then by dressing $|\uparrow\rangle$ with a small component of the $5P_{3/2}$ excited state, the effective linewidth of $|\uparrow\rangle$ can be controllably set to, for example, 100 kHz.

\section{Keldysh calculations in the Gaussian approximation}

In what follows we briefly describe how excitation gaps, fluctuations, and squeezing can be computed using the Keldysh technique~\cite{kamenev:review, dallatorre2013}. We shall analyze the following master equation:
\beq
\dot{\rho} = -i [H, \rho] + \gamma \sum_i \left[b_i \rho b^\dagger_i - \frac{1}{2} (\rho b^\dagger_i b_i + b^\dagger_i b_i \rho)\right].
\eeq
For notational convenience, we choose dimension $d=1$. Then
\beq
H =  \frac{J_x + J_y}{2} \sum_{\langle ij \rangle} (b^\dagger_i b_j + b_i b_j^\dagger) + \frac{J_x - J_y}{2} \sum_{\langle ij \rangle} (b^\dagger_i b^\dagger_j + b_i b_j) - 2 J_z \sum_i b^\dagger_i b_i.
\eeq
In what follows we define $f \equiv J_x + J_y$ and $\Delta \equiv J_x - J_y$. Using standard identities, one can rewrite the master equation as the following Keldysh action, in Fourier space:
\bea
S & = & \int d\omega dk \Bigg\{ \sum_{\alpha = \pm} \alpha \Big[  [\omega + 2 J_z - f \cos(k)] [b_\alpha^*(\omega,k) b_\alpha(\omega, k) + b_\alpha^*(-\omega,-k) b_\alpha(-\omega, -k)] \\
&& \qquad \qquad \qquad - \Delta \cos(k)[b_\alpha^*(\omega, k) b_\alpha^*(-\omega, -k) + b_\alpha(\omega, k) b_\alpha(-\omega, -k)] \Big] \nonumber \\
& & \qquad \qquad - i \gamma \left[ b_+(\omega, k) b_-^*(\omega, k) - \frac{1}{2} [ b_+(\omega, k) b_+^*(\omega, k) + b_-(\omega, k) b_-^*(\omega, k) ] \right]\\
&& \qquad \qquad - i \gamma \left[ b_+(-\omega, -k) b_-^*(-\omega, -k) - \frac{1}{2} [ b_+(-\omega, -k) b_+^*(-\omega, -k) + b_-(-\omega, -k) b_-^*(-\omega, -k) ] \right] \Bigg\},
\eea
where $b_\alpha(\omega, k)$ is now a complex field.
If we define the fields $b_{cl} = (b_+ + b_-)/\sqrt{2}$ and $b_q = (b_+ - b_-)/\sqrt{2}$ (where all other arguments are assumed to be the same), then this action can be written in the compact $4 \times 4$ form:
\beq
S = \int d\omega dk \left( \begin{array}{cccc} b^*_{cl} (\omega, k) \quad&\quad b_{cl} (-\omega, -k) \quad&\quad b^*_{q} (\omega, k) \quad&\quad b_{q} (-\omega, -k) \end{array} \right) \left( \begin{array}{cc} 0 & D^{R\dagger} \\ D^R & D^K  \end{array} \right) \left( \begin{array}{c} b_{cl}(\omega, k) \\ b^*_{cl}(-\omega, -k) \\ b_q(\omega, k) \\ b_q^*(-\omega, -k) \end{array} \right)
\eeq 
in terms of the inverse retarded Green's function
%
%
\beq
D^R = \left(\begin{array}{cc} \omega - f \cos k + 2J_z + i\gamma/2 & - \Delta \cos k \\ - \Delta \cos k & -\omega - f \cos k + 2J_z - i\gamma/2 \end{array} \right)
\eeq
and the inverse Keldysh Green's function
\beq
D^K = \left( \begin{array}{cc} i \gamma & 0 \\ 0 & i \gamma \end{array} \right).
\eeq

\subsection*{Relaxation rate}

In order to find the relaxation rate (i.e., the eigenvalue gap of the Liouvillian) we use the fact that the characteristic frequencies in the system are given by the zeros of the determinant of $D^R$. These frequencies correspond to the poles of the retarded Green's function ~\cite{kamenev:review}, and are easily computed to be
\beq
\omega = - i \gamma/2 \pm \sqrt{(f \cos k - 2 J_z)^2 - \Delta^2 \cos^2 k}.
\eeq
When the $\omega$ vanishes, there is a second-order phase transition. This occurs when $\Delta^2 \cos^2 k - (f \cos k - 2 J_z)^2 = \gamma^2 / 4$, which, when re-expressed in terms of $J_x$ and $J_y$, gives the result shown in the main text.

\subsection*{Expectation values and squeezing}

In order to compute expectation values and correlations in the steady state, it is necessary to compute the so-called Keldysh Green's function $G_K$~\cite{kamenev:review, dallatorre2013}, which is given by the matrix product
\beq
G_K = - (D^R)^{-1} D^K (D^{R\dagger})^{-1}.
\eeq
We consider the case of $J_z=0$. Transforming to the time domain and setting $t=0$, we find that
\bea
i G_K(k) &  \equiv & \left( \begin{array}{cc} 2 \langle b^\dagger(k) b(k) \rangle + 1 & 2 \langle b^\dagger(k) b^\dagger(-k) \rangle \\ 2 \langle b(k) b(-k) \rangle &  2 \langle b^\dagger(k) b(k) \rangle + 1 \end{array} \right) \nonumber \\ 
& = & \frac{1}{\gamma^2+4(f^2-\Delta^2)\cos^2 k} \left( \begin{array}{cc} \gamma^2 + 4f^2 \cos^2 k & - 2\Delta \cos k( 2f\cos k + i \gamma) \\ - 2\Delta \cos k( 2f\cos k - i \gamma) & \gamma^2 + 4f^2 \cos^2 k  \end{array} \right). 
\eea
From this, it is straightforward to arrive at the expression for $\langle b^\dagger b \rangle$ given in the main text. Moreover, using the definition of squeezing in Ref.~\cite{ma2011}, one can compute the squeezing as plotted in the main text.

\section{Liouvillian gap}
Below are plots of the Liouvillian gap, calculated numerically. The Liouvillian gap is defined as the real part of the eigenvalue of the master equation with largest nonzero real part. It corresponds to the rate of approach to the steady state. The gap was calculated using exact diagonalization for a 1D chain of 6 spins with nearest-neighbor interactions. The gap become small at the boundary of the paramagnetic phase (compare with the phase diagram in the main text), indicating a critical slowing down. In particular, for the case of $J_z/\gamma=1$, gap clearly shows the transition from the paramagnetic phase to the ferromagnetic and antiferromagnetic phases. The gap also becomes small at the transition from the paramagnetic phase to the spin-density-wave phase (the oblong shape is probably due to the small size of the system, since only certain wave vectors are allowed).

For an infinite system in two or more dimensions, we expect the gap to vanish along the boundary of the paramagnetic phase; the fact that the gap does not vanish here is due to the finite size and the low dimensionality. Note that the gap is 1/2 along the $J_x=J_y$ line, because the Hamiltonian contains only flip-flop terms $(\sigma_m^+ \sigma_n^- + \sigma_m^-\sigma_n^+)$ that do not counteract the overall decay.

\begin{figure}[h]
\centering
\includegraphics[width=7in,clip]{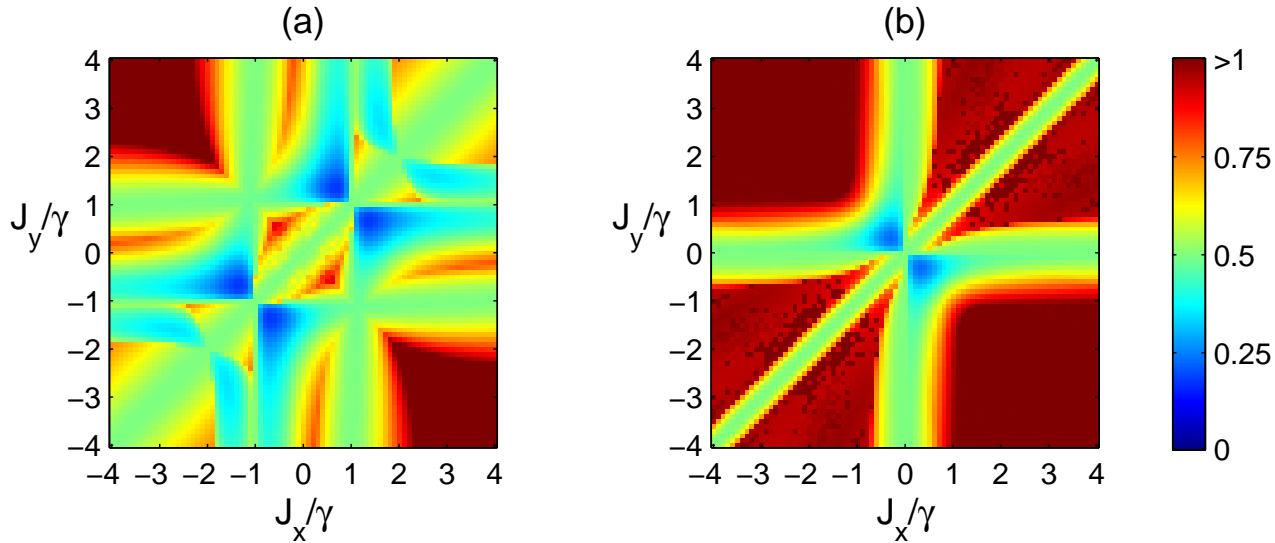}
\caption{\label{fig:gap} Gap of the Liouvillian for (a) $J_z/\gamma=1$ and (b) $J_z=0$.}
\end{figure}

\end{widetext}

\end{document}